\documentclass[final,5p,times,twocolum,authoryear]{elsarticle}

\usepackage{amssymb}
\usepackage{lipsum}
\usepackage{amsthm}

\journal{New Astronomy}

\begin{document}

\begin{frontmatter}

\title{The First Light Curve Analysis of Twin Binary System V1175 Cas using Ground-based and TESS data}

%\tnotetext[mytitlenote]{Fully documented templates are available in the elsarticle package on 
%\href{http://www.ctan.org/tex-archive/macros/latex/contrib/elsarticle}{CTAN}.}

\author{Neslihan Alan}
\affiliation{organization={Istanbul University  Faculty of Science, Department of Astronomy and Space Sciences},%Department and Organization
            addressline={Istanbul University, Faculty of Science, Department of Astronomy and Space Sciences, 34119, Beyazit, Istanbul, Turkey}, 
            city={Istanbul},
            postcode={34119}, 
            state={},
            country={Turkey}}

\ead{neslihan.alan@gmail.com}

\begin{abstract}
Eclipsing binary systems hold a central position within astrophysics in that the fundamental parameters of stars can be determined by direct observations. The simultaneous analyses of high-quality space observations, combined with ground-based photometric data, have allowed more sensitive detection of fundamental stellar parameters by multicolor photometry. In the paper, the fundamental parameters of the component stars for the V1175 Cas binary system were sensitively obtained by a simultaneous analysis of the Transiting Exoplanet Survey Satellite ({\it TESS}) light curve, and new CCD observations in {\it BVRI} filters obtained with a 60 cm robotic telescope at the T\"UB\.ITAK National Observatory. Following the analysis, the masses and radii of the primary and secondary binary components were determined as $M_{1}= 1.64\pm 0.04\,M_\odot$, $M_{2}= 1.63\pm0.07\,M_\odot$, and $R_{1}=1.77\pm 0.05\,R_\odot$, $R_{2}= 1.77\pm 0.25\,R_\odot$, respectively. Moreover, the distance of V1175 Cas was computed as $280\pm32$ pc. The photometric analysis reveals that the components of the system are in a similar evolutionary state. The primary and secondary components exhibit nearly the same masses, while their radii are perfectly matched. Additionally, the ages of the components are also consistent within the statistical uncertainties. Consequently, the system's overall age is assessed to be approximately $750\pm70$ Myr.
\end{abstract}

\begin{keyword}
\texttt{}techniques: photometric: --- Stars; binaries; eclipsing, Stars; fundamental parameters, Stars; evolution --- stars: individual: V1175 Cas
%\MSC[2010] 00-01\sep  99-00
\end{keyword}
\end{frontmatter}
%\linenumbers

\section{Introduction}           %% first-level sections will be auto-capitalized
\label{sect:intro}

Eclipsing binary stars are considered invaluable laboratories for astronomical research, playing a vital role in understanding stellar structure and evolution, as well as galaxy dynamics. The fundamental stellar parameters such as mass ($M$), radius ($R$), and luminosity ($L$) can be obtained directly from their observations. These systems offer a unique chance to accurately determine these essential parameters. The fundamental stellar parameters can be obtained more precisely, especially by using the high-quality photometric data of space telescopes such as {\it TESS} \citep{Ricker15}. Enhanced sensitivity in the fundamental parameters of stars enables theoretical models to be checked and evolutionary models to be compared with observational findings to create more realistic models. Detached eclipsing binary systems are particularly suitable for this goal because of the relatively weak interactions between the components. The evolutionary models of single stars can be investigated with fundamental stellar parameters calculated from observations of detached eclipsing binaries, and the consistency of theoretical evolutionary models with observations can be meticulously tested. Furthermore, an equally important aspect is the introduction of fundamental stellar parameters into the literature through detailed light curve analyses of detached eclipsing binaries, which have never been investigated in detail before. To address this, turn attention to the case of V1175 Cas, a detached binary system for which only minima times have been reported and no light curve analysis has been performed until now. 

V1175 Cas is classified as an Algol-type eclipsing binary system according to \citet{Kazarovets11}. Despite this classification, an exhaustive investigation of the system has remained absent thus far, and the fundamental stellar parameters of V1175 Cas remain a mystery. To unravel these parameters, an essential approach involves a thorough analysis of the system's light curve. With this objective in mind, this study undertakes the first light curve analysis of V1175 Cas, employing data from {\it TESS} and T60 observations. Through this analysis, the fundamental parameters of the system were derived. Table 1 lists general catalogue information about the V1175 Cas.

\label{sect:general inf}
\begin{table}
  \caption{Catalogue information of V1175 Cas}
  \begin{center} \label{rv_table}
           \begin{tabular}{lcc}
\hline
${\rm RA^1}$ & $03^{\rm h} 21^{\rm m} 26^{\rm s}.53$\\
${\rm DEC^1}$ & $+73^{\circ} 26^{'} 07^{''}.93$\\
${\rm Type^2}$ & EA\\  
Parallax ${\rm (mas)^1}$ &3.60 \\
$V$ magnitude ${\rm (mag)^1}$ & $9.52$ \\
$P$ ${\rm (day)^2}$ & 3.46 \\
 \hline
     \end{tabular}\\
$^1$ refers to the data taken from SIMBAD\footnote{https://simbad.cds.unistra.fr/simbad/sim-fbasic}, $^2$ represents the O-C Gateway\footnote{http://var2.astro.cz/ocgate/} data. 
     \end{center}
\end{table}

The structure of the paper that presents the first light curve analysis findings for V1175 Cas is briefly described in the following: In Section 2, details about the observational data and an outline of the methodology for calculating the new light elements are given. Moving on to Section 3, the process of conducting a simultaneous analysis of the {\it TESS} data alongside ground-based photometric data is elucidated. The results of this analysis, along with the derivation of the fundamental parameters of V1175 Cas, are elaborated upon in Section 4. Finally, Section 5 presents a comprehensive discussion of the results and the state of evolution of the system.

\section{Observational data}
\label{sect:Obs}

The new multicolor CCD observations of V1175 Cas were performed over 117 nights between September 2018 and October 2019 with the 60~cm robotic telescope (T60) at the T\"UB\.ITAK National Observatory (TUG). Controlling the T60 is achieved through the implementation of the OCAAS open-source software, formally named TALON (see \citet{Parmaksizoglu14}). This telescope is equipped with the FLI ProLine 3041-UV CCD until the 23rd of July, 2019. Subsequently, a transition was made to the Andor iKon-L 936 BEX2-DD model camera for operations after that particular date. The FLI ProLine 3041-UV CCD offers an image scale where each pixel corresponds to 0.51 arcsec, providing a field of observation (FOV) of 17.4 arcmin. The Andor iKon-L 936 BEX2-DD camera, on the other hand, provides a finer image scale of 0.456 arcsec per pixel, giving an observation FOV of 15.6 arcmin.

In the observations of V1175 Cas, Bessell {\it BVRI} filters were used (see \citet{Bessell90}). The exposure time was set to 5 s, the same for all $BVRI$ filters. Calibration images, including sky flats and bias frames, were taken at intervals throughout the observations to correct for pixel-to-pixel variations on the CCD chip. For comparison and confirmation purposes, TYC 4338-973-1 and TYC 4338-1025-1 were used as comparison and check stars, respectively. In addition to ground-based observations, light curve analysis involved using data from the {\it TESS}. {\it TESS} extensively scans most of the entire sky in sectors, with a dedicated observing time of 27.4 days per sector.  Operating within the 600-1000 nm wavelength range, {\it TESS} provides broadband photometric data \citep{Ricker15}. The V1175 Cas {\it TESS} data from sector 19 were acquired between November 28th and December 23rd, 2019, utilizing a 120-second exposure time. {\it TESS} data of the system were sourced from the Mikulski Archive for Space Telescopes (MAST)\footnote{https://archive.stsci.edu/} database. For the analysis, Pre-search Data Conditioning Simple Aperture Photometry light curves, as introduced by \citet{Ricker15}, were employed. The photometric data exhibited an average error of approximately 0.1\%. 

The data reduction of the T60 observations consists of several steps. Initially, bias and dark frames were subtracted from the science frames, followed by a flat-fielding correction. Subsequently, the resulting reduced CCD images were utilized to compute the differential magnitudes of the target stars. This particular procedure was accomplished using MYRaf software, the IRAF aperture photometry GUI tool developed by \citet{Kilic16}. Notably, no significant light variations were detected in the comparison and check stars throughout the observations. The external uncertainties for comparison minus check magnitudes were quantified to be about 29 mmag in $B$, 24 mmag in $V$, 20 mmag in $R$, and 20 mmag in $I$ filters. These values were derived from the standard deviation of differential magnitude variation between the comparison and check stars during the same night. Observational data were not converted to the standard Bessell {\it BVRI} system and differential magnitudes were used in light curve analyses.

The minima times of the system were determined using {\it TESS} data. The primary minima times were procured from the first, middle, and last parts of {\it TESS} sector 19 observations. Meanwhile, the secondary minima time was derived from the middle part of the {\it TESS} dataset. In total, four {\it TESS} minima times were obtained. The literature minima times of V1175 Cas were also collected from the O-C Gateway. We used 17 minima times and investigated the period changes of the system. Considering the first The All Sky Automated Survey \citep[ASAS,][]{Paczynski06} minima in the O-C Gateway with all the minima times, a parabolic fit for the period change could be obtained. However, due to the fewer minima times, this result was not reliable and the system needs more minima times to obtain more decent results. Therefore, in this study, the first ASAS minima time was ignored, and only determined new light elements of V1175 Cas by finding a linear fit to the other 16 minima times. The new light elements are given in the following equation:

Ephemeris
\begin{eqnarray}\label{eq1}
{\rm HJD(MinI)}=2458817.7506(9)+3^{\rm d}.457179 (2)\times E
\end{eqnarray}

The values given inside the parentheses in the equation refer to the errors at the last digit for the light elements.

\section{Light Curve Modelling}
\label{sect:LightCurve}

A comprehensive analysis of the light curve data for V1175 Cas using various photometric filters, including normalized $BVRI$ and {\it TESS} data were performed in this work. The Wilson-Devinney (W-D) \citep{WilsonDevinney71} code, integrated with a Monte Carlo simulation \citep{Zola04, Zola10}, was employed for this analysis to precisely estimate uncertainties in the parameters being adjusted. The approach employed for analysis with the W-D code involved a combination of fixed parameters based on theoretical models and prior research, as well as parameters that are iteratively adjusted through successive iterations. The fixed and adopted parameters are defined in the following. For the V1175 Cas system, an unreddened color index of $(B-V)_{\rm 0}=0.296\pm0.02$ mag was derived using $E_{\rm d}(B-V)=0^{\rm m}.174$ calculated according to the \citet{Schlafly11} calibration and $B-V = 0^{\rm m}.399\pm0.02$ from the Tycho-2 catalogue \citep{Hog00}. The initial temperature of the primary component of the system was fixed at $T_{\rm 1,eff}=7150$ K corresponding to the $B-V$ color index using the astrophysical parameters of main sequence stars presented by \citet{Eker20}. Simultaneous analysis of ground-based $BVRI$ light curves and {\it TESS} data was performed, with mode 2 used for detached binary systems due to the nature of light variations observed. The root square limb darkening law was adopted, and the limb darkening coefficients were taken from the \citet{Vanhamme93} tables, based on the filter wavelengths and temperatures of the components of V1175 Cas. It's worth noting that {\it TESS} passband was not directly included in the W-D code, therefore, the $I$-band was used for binary modeling as {\it TESS} passband aligns with Cousins $I$-band. The constant coefficients were also selected taking into account the $I$-band. With the convective atmosphere ($T_{\rm eff}<7200$ K) approximation, the bolometric gravity-darkening exponents of the components were taken as 0.32 from \citet{Lucy67}, retaining their bolometric albedo fixed at 0.5 following \citet{Rucinski69}. Both components were assumed to be in synchronous rotation ($F_{\rm 1}=F_{\rm 2}=1$). The secondary minima of V1175 Cas is in phase 0.5 and there is no noticeable asymmetry in the observed light curve. Moreover, the ascent and the descent durations were the same for the primary and secondary minima. This led to the assumption of a circular orbit ($e=0$). The remaining parameters, including the orbital inclination ($i$), the surface temperature of secondary ($T_{\rm 2,eff}$), the dimension-less surface potential of primary and secondary components ($\Omega_{1,2}$), phase shift, the mass ratio ($q$), and fractional luminosity of primary component ($L_{1}$), were considered as adjustable parameters. The presence of a third body contributing to the total light was assessed by considering a free parameter $l_{3}$ in the solution. However, V1175 Cas was found to have negligible third light effects, and thus, $l_{3}$ was not included in the final solution. The final model parameters are detailed in Table 2, and the comparison between observed and computed light curves is illustrated in Fig. 1. Additionally, the Roche geometry of the system is depicted in Fig. 2.

\begin{table}
\begin{center}
\centering
\label{lcresult}
\caption{Findings of the light curve analysis of V1175 Cas. The subscripts 1, 2 and 3 represent the primary, the secondary, and third components, respectively. $^a$ shows the fixed parameters.}
\begin{tabular}{lrrr}
\hline
 Parameter			 			               & Value		\\	
\hline
$T_{0}$ {\rm (HJD+2400000)}                     & 58817.7506 \\
$P_{\rm orb}$ (days)                            & 3.457179  \\
$i$ ($^{\rm o}$)	       	                    & 86.034 $\pm$ 0.012  \\	
$T$$_{\rm 1,eff}$$^a$ (K)                 		        & 7150 $\pm$ 150	\\	
$T$$_{\rm 2,eff}$ (K)    	          		            & 7090 $\pm$ 156		\\
$e$	         	               		            & 0.000		  \\
$\Omega$$_{1}$		           	                & 9.062 $\pm$ 0.033  	  \\
$\Omega$$_{2}$		            	            & 9.012 $\pm$ 0.296  	  \\
Phase shift             	  	                & 0.0008 $\pm$ 0.0001	  \\
$q$                     	  	                & 0.995 $\pm$ 0.032  	  \\
$r$$_{\rm 1}^*$ (mean)                          & 0.1242 $\pm$ 0.0007 \\
$r$$_{\rm 2}^*$ (mean)                          & 0.1244 $\pm$ 0.0036 \\
$L$$_{\rm 1}$/($L$$_{1}$+$L$$_{2}$) ($TESS$) 	& 0.509 $\pm$ 0.001   \\
$L$$_{\rm 1}$/($L$$_{1}$+$L$$_{2}$) ($B$)   	& 0.508 $\pm$ 0.006   \\
$L$$_{\rm 1}$/($L$$_{1}$+$L$$_{2}$) ($V$)   	& 0.507 $\pm$ 0.006   \\
$L$$_{\rm 1}$/($L$$_{1}$+$L$$_{2}$) ($R$)  	    & 0.505 $\pm$ 0.005   \\
$L$$_{\rm 1}$/($L$$_{1}$+$L$$_{2}$) ($I$)   	& 0.505 $\pm$ 0.005   \\
$L$$_{\rm 2}$/($L$$_{1}$+$L$$_{2}$) ($TESS$)    & 0.491 $\pm$ 0.002 	 \\
$L$$_{\rm 2}$/($L$$_{1}$+$L$$_{2}$) ($B$)   	& 0.492 $\pm$ 0.006 	 \\
$L$$_{\rm 2}$/($L$$_{1}$+$L$$_{2}$) ($V$)  	    & 0.493 $\pm$ 0.006 	 \\
$L$$_{\rm 2}$/($L$$_{1}$+$L$$_{2}$) ($R$)   	& 0.495 $\pm$ 0.006 	 \\
$L$$_{\rm 2}$/($L$$_{1}$+$L$$_{2}$) ($I$)   	& 0.495 $\pm$ 0.005 	 \\
$l$$_{\rm 3}$ ($TESS$)                        	& 0.0	        	  \\
 \hline
\end{tabular}
     \end{center}
     \begin{description}
     \centering
 \item[ ] * fractional radii.
 \end{description}
\end{table}

\begin{figure}
\begin{center}
\includegraphics[width=\columnwidth]{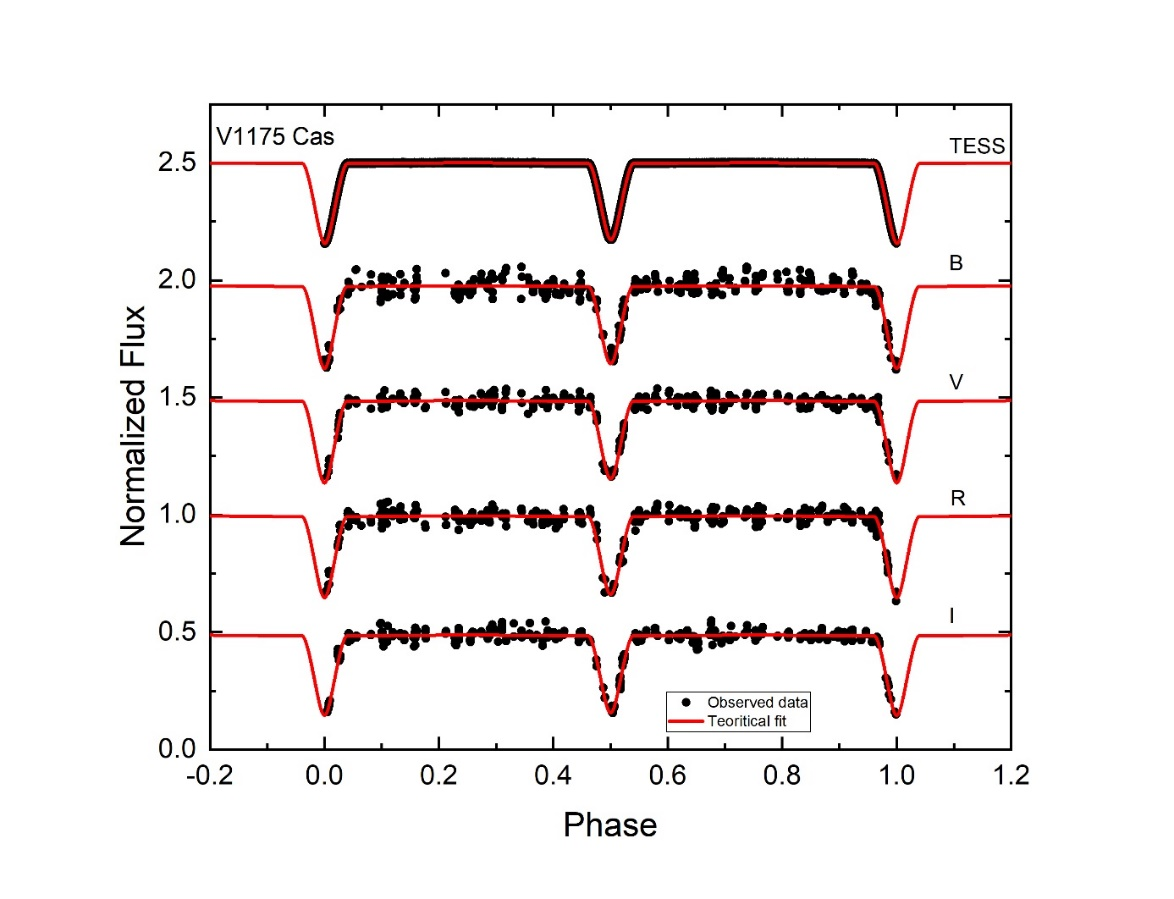}
\caption{Comparison of the observational data (black dot) with the theoretical light curves (red line) of V1175 Cas. } \label{LC_fit}
\end{center}
\end{figure}

\begin{figure}
\begin{center}
\includegraphics[width=0.9\columnwidth]{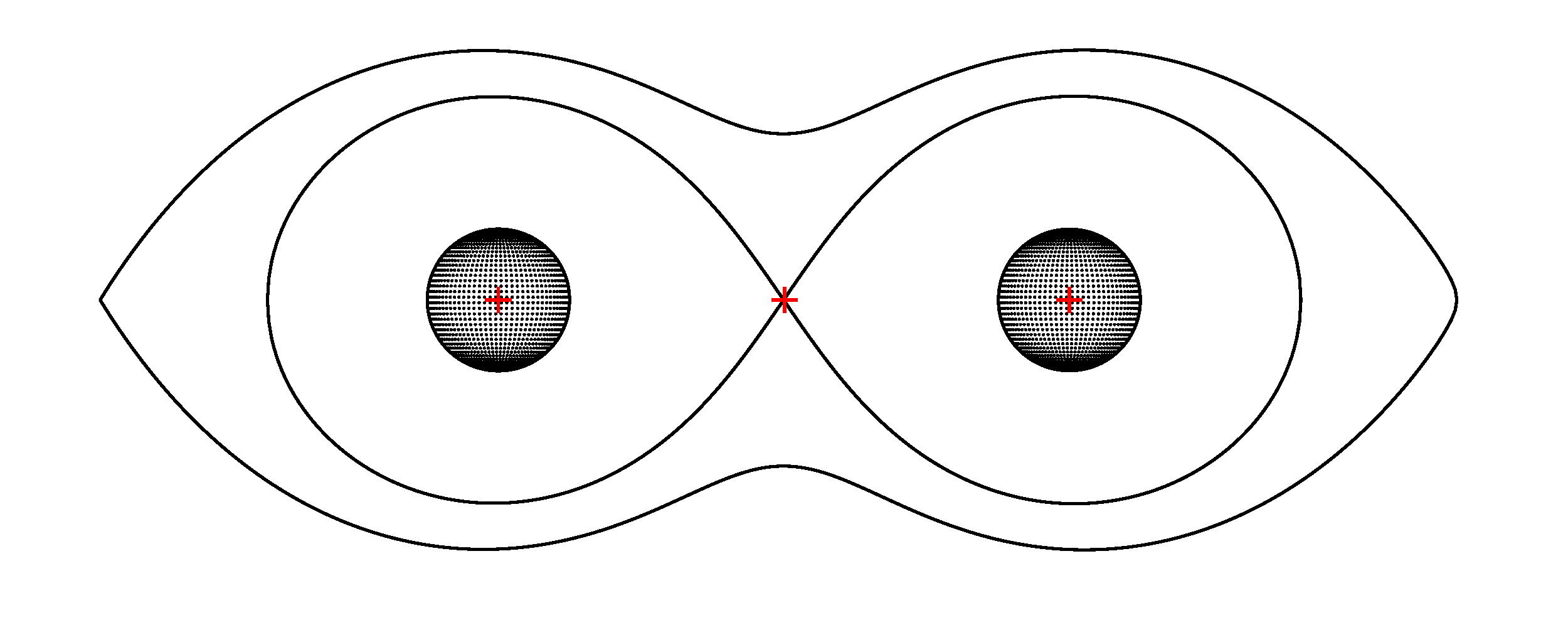}
\caption{Roche geometry of V1175 Cas obtained using the parameters of the best light curve model.} \label{LC_fit}
\end{center}
\end{figure}

\section{Estimated Fundamental Parameters and Evolutionary Status}
\label{sect:Absolute Parameters}

As no spectral observations of the V1175 Cas detached binary system have been performed, the radial velocity curves for its components are non-existent. Nevertheless, it is possible to roughly estimate the fundamental parameters for these components based on information obtained from light curve analysis. The fundamental parameters are listed in detail in Table 3. By assuming the primary component to be a main sequence star characterized by an effective temperature $T_{\rm 1,eff}=7150$ K, the mass of the primary component was designated to be $M_{1}=1.64\,M_{\odot}$ from the correlation between mass and $T_{\rm eff}$ for main sequence stars given by \citet{Eker20}, for V1175 Cas. The mass of the secondary component was computed from the mass ratio obtained from the photometric light curve analysis. Using the fractional radii from Table 2 and the semi-major axis calculated via Kepler's third law, the radii of the components were ascertained. By adopting standard solar values ($T_{{\rm eff, \odot}} = 5777$ K, $M_{{\rm bol},\odot} = 4^{\rm m}.74$) and employing bolometric corrections from \citet{Eker20}, the luminosities and bolometric magnitudes of the components were estimated. Surface gravity values were determined for the primary and secondary components respectively as follows: log\,$g_{1} = 4.16 \pm 0.03$ and log\,$g_{2}=4.15 \pm 0.12$ in cgs units for V1175 Cas. This finding points to the almost same evolutionary status of each of the components. 

Utilizing reddening maps from \citet{Schlafly11} and the NASA Extragalactic Database\footnote{https://ned.ipac.caltech.edu}, a value of $E_{\rm d}(B-V)$ of about $0^{\rm m}.174$ was calculated for the Galactic coordinates ($l = 133^{\circ}.22$, $b =+13^{\circ}.64$) of the system. Correspondingly, the interstellar visual extinction ($A_{\rm V}$) is found as $0.539\pm 0.029$ mag for V1175 Cas, using the common formula $A_{\rm V}= 3.1 \times E_{\rm d}(B-V)$ \citep[for details of the method, see][]{Bilir08, Eker09}. Based on the interstellar extinction given in Table 3, the apparent magnitude of the system, component light ratios listed in Table 2, along with the $BC_{1}$= $0^{\rm m}.077$ and $BC_{2}$= $0^{\rm m}.078$ values calculated according to \citet{Eker20}), the distance of V1175 Cas was derived as $280\pm32$ pc. The accurate fundamental stellar parameters obtained provide a comprehensive insight into the binary component stars' evolutionary status and age. The examination of these parameters led us to utilize the MESA Isochrones \& Stellar Tracks (MIST) framework \citep{{Choi16},{Dotter16},{Paxton11},{Paxton13},{Paxton15},{Paxton18}} to delve into the evolutionary status of the components. Notably, the best theoretical fit to the calculated fundamental parameters in the Hertzsprung Russell diagram was found to be the evolutionary track characterized by $Z$\,=\,0.020 $\pm$ 0.002 (depicted in Fig. 3). Further insights were gained through analysis of the $\log R-age$ diagrams, allowing us to deduce that the age of the binary system is approximately 750\,$\pm$\,70 Myr (illustrated in Fig. 4).

\begin{table}
\begin{center}
\centering
\label{lcresult}
\caption{Estimated fundamental stellar parameters of V1175 Cas}
\begin{tabular}{lrr}
\hline
  Parameter 			              & Value			\\	
\hline
$M$$_{1}$ ($M_\odot$)	          	&1.64 $\pm$ 0.04   \\	
$M$$_{2}$ ($M_\odot$)	         	&1.63 $\pm$ 0.07   \\
$R$$_{1}$ ($R_\odot$)	          	&1.77 $\pm$ 0.05   \\
$R$$_{2}$ ($R_\odot$)		  		&1.77 $\pm$ 0.25   \\
$a$ ($R$$_{\odot}$)        	   	    &14.27$\pm$ 0.18   \\
$\log L$$_{1}$ ($L_\odot$)		  	&0.87 $\pm$ 0.05   \\
$\log L$$_{2}$ ($L_\odot$)		  	&0.86 $\pm$ 0.15   \\
$\log g$\,$_{1}$ (cgs)              &4.16 $\pm$ 0.03   \\
$\log g$\,$_{2}$ (cgs)              &4.15 $\pm$ 0.12   \\
$M_{\rm Bol, 1}$ (mag)              &2.57 $\pm$ 0.25   \\
$M_{\rm Bol, 2}$ (mag)	 	        &2.60 $\pm$ 0.58   \\
$M_{\rm V, 1}$ (mag)	         	&2.49 $\pm$ 0.25   \\
$M_{\rm V, 2}$ (mag)	          	&2.52 $\pm$ 0.44   \\
$A_{\rm V,d}$ (mag)	         	    &0.539 $\pm$ 0.029 \\
$d$ (pc)                            &280 $\pm$ 32  	   \\
 \hline
\end{tabular}
     \end{center}
\end{table}

\begin{figure}[!ht]
\begin{center}
\includegraphics[width=\columnwidth]{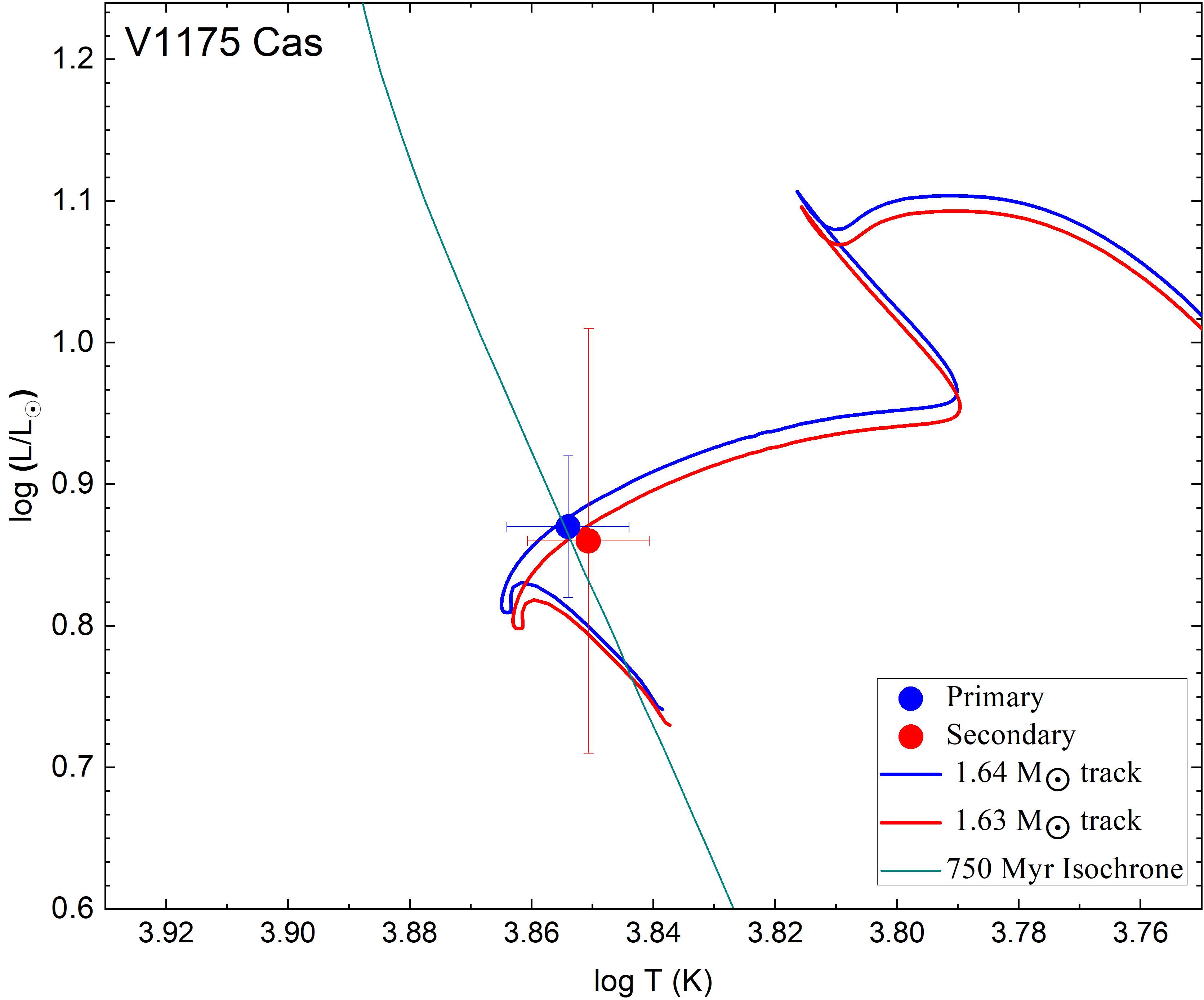}
\caption{In the log\,$L$ - log\,$T$ plane, the positions of the primary (blue dot) and secondary (red dot) components of V1175 Cas. Evolutionary tracks according to the metallicity value of $Z = 0.020$ are depicted by blue and red lines for the primary and secondary component stars, respectively. Theoretical evolutionary curves were computed by MESA code.} \label{LogL_Te}
\end{center}
\end{figure}

\begin{figure}[!ht]
\begin{center}
\includegraphics[width=\columnwidth]{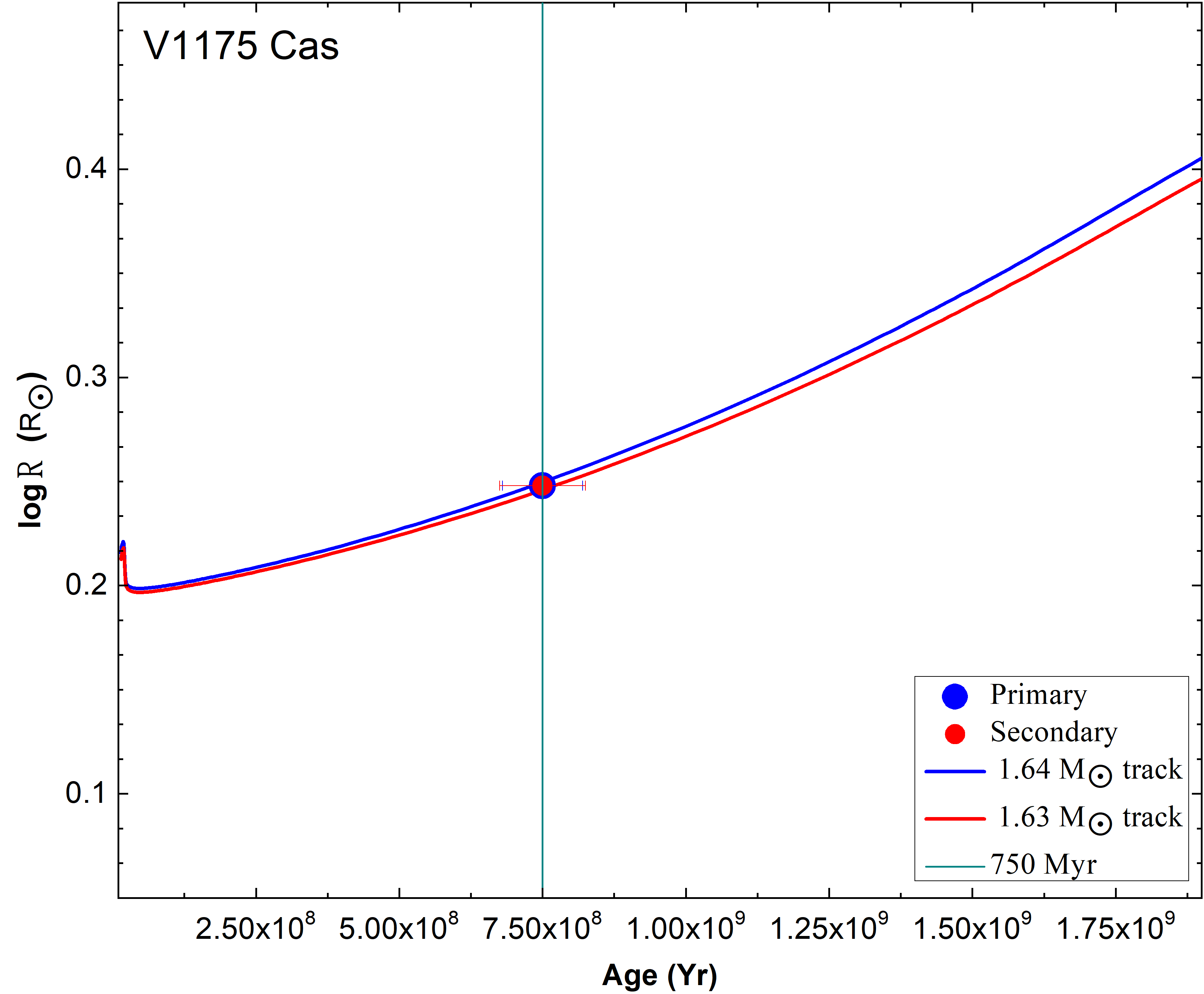}
\caption{In the plane of log\,$R$ - Age, positions of the primary (blue dot) and secondary (red dot) components of V1175 Cas. Evolutionary tracks according to metallicity value of $Z=0.020$ are represented by blue and red lines for the primary and secondary component stars, respectively.} \label{R_age}
\end{center}
\end{figure}

\section{Discussion and Conclusions}
\label{sect:discussion}

This paper presents the first light curve analysis of the high-quality {\it TESS} data and CCD multicolor observational datasets in {\it BVRI} filters to derive the fundamental parameters of the components of V1175 Cas. Furthermore, new light elements of the system were obtained by using the minima times calculated from {\it TESS} data and those collected from the literature. Consequent to the light curve analysis, the respective masses of the primary and secondary components were determined as $M_{1}$\,=\,1.64\,$\pm$\,0.04\,$M_\odot$ and $M_{2}$\,=\,1.63\,$\pm$\,0.07\,$M_\odot$. Regarding the primary and secondary components of the system, the radii were found as follows: be $R_{1}$\,=\,1.77\,$\pm$\,0.05\,$R_\odot$ and $R_{2}$\,=\,1.77\,$\pm$\,0.25\,$R_\odot$, respectively. In addition, the light curve analysis indicates that the distance of V1175 Cas is approximately 280 $\pm$ 32 pc, a value clearly in accordance with the {\it Gaia}-DR3 distance of 278 $\pm$ 1 pc \citep{Gaia23}. The temperatures of the primary and secondary components of V1175 Cas were obtained by photometric analysis of the system as $T_{\rm 1,eff}\,=\,7150$ K and $T_{\rm 2,eff}\,=\,7090$ K, respectively. By comparing the achieved temperatures with the temperatures in the table reported by \citep{Eker20} for main sequence stars, the spectral types of both components were found to be F0. No detailed photometric or spectral analysis of the system has been performed before, and therefore there is no information on the spectral types of the component stars in the literature, which is provided by this work. 

\citet{LucyRicco79} listed the twin binaries according to their mass ratio and found a peak at $q$ $\approx$ 0.97. \citet{Griffin85} also reported that double-lined spectroscopic binaries on the main sequence show a net peak around $q$ $\approx$ 1. The mass ratio of V1175 Cas derived in this work is $q=0.995\pm0.032$ which is in strong agreement with values given by \citet{LucyRicco79} and \citet{Griffin85}. Additionally, the calculated $q$ value is consistent, within statistical uncertainties, with the median mass ratio value $q=0.931$ given by \citet{Eker14} for detached binary systems with the primary component of spectral type F. The mass values calculated for each component were substituted into the mass-luminosity correlation in the corresponding mass range given by \citet{Eker18} for main sequence stars. Based on this correlation, the mean luminosity values for the primary and secondary components were calculated as $\log L_1=0.94\,L_{\odot}$ and $\log L_2=0.93\,L_{\odot}$, respectively. It was found that the mean luminosity values calculated from the \citet{Eker18} equations were slightly larger than the observational luminosity values ($\Delta L$ = 0.07$\,L_{\odot}$). This shows that the components of V1175 Cas have lower luminosity values than the stars of \citet{Eker18} in the $1.05<M/M_\odot<2.40$ mass range, but it also indicates that the age of the system is younger than the average age of the main sequence. Indeed, the average age of \citet{Eker18} in the $1.05<M/M_\odot<2.40$ mass range is 1.92 Gyr, which is considerably older than the 0.75 Gyr age of V1175 Cas. 
 
The components of the system are on the main sequence and have similar evolutionary status as anticipated. The masses of the primary and secondary components are quite close and the radii are exactly the same. The ages of the components are also consistent within the statistical uncertainties. The age of the system is estimated at 750 $\pm$ 70 Myr. Since V1175 Cas is composed of close-mass components, observing the system's spectral characteristics is necessary to understand the variations arising from small mass differences during their evolutionary processes and to engage in more detailed discussions about their evolution \citep[e.g.][]{Alicavus22}. Therefore, this system is significant for future studies involving evolutionary modeling.

In this research, the components of the V1175 Cas system were investigated using highly sensitive photometric data. This allowed to determine the fundamental stellar parameters for the components. Detached binary systems such as V1175 Cas provide a unique opportunity to directly measure stellar masses, radii, and luminosities, which are difficult to accurately quantify for single stars. Collectively, the V1175 Cas binary system provides a precious laboratory for extending our understanding of stellar evolution, binary star interactions, and astrophysical phenomena. To improve the accuracy of calculating component mass ratios and their fundamental stellar parameters, acquiring the radial velocity curves of the components is crucial. This necessitates performing spectroscopic observations on V1175 Cas. By combining spectroscopic data with photometric information in the light curve analysis, future observations can provide even more precise findings. The simultaneous analysis of spectroscopic and photometric data from the system has the potential to yield significantly refined results.

\section*{Acknowledgements}
Many heartfelt thanks to the referee for insightful and constructive suggestions that significantly improved the paper. This study was funded by the Scientific Research Projects Coordination Unit of Istanbul University. Project number: 37903. The author would like to thank T\"UB\.ITAK National Observatory (TUG) for partial support towards using the T60 telescope via project 18BT60-1324. The author also thanks the on-duty observers and the technical staff at the TUG for their support before and during the observations. A special thanks to Fahri Ali\c{c}avu\c{s} and Mehmet Alpsoy for their valuable suggestions and contributions, and Sel\c{c}uk Bilir for the inspiration and helpful discussions.

This research has made use of NASA’s (National Aeronautics and Space Administration) Astrophysics Data System and the SIMBAD Astronomical Database, operated at CDS, Strasbourg, France, and NASA/IPAC Infrared Science Archive, which is operated by the Jet Propulsion Laboratory, California Institute of Technology, under contract with the National Aeronautics and Space Administration. This work also has made use of data from the European Space Agency (ESA) mission $Gaia$\footnote{https://www.cosmos.esa.int/gaia}, processed by the $Gaia$ Data Processing and Analysis Consortium (DPAC)\footnote{https://www.cosmos.esa.int/web/gaia/dpac/consortium}. 
Funding for the DPAC has been provided by national institutions, in particular, the institutions participating in the Gaia Multilateral Agreement. The {\it TESS} data presented in this paper were obtained from the Mikulski Archive for Space Telescopes (MAST). Funding for the {\it TESS} mission is provided by the NASA Explorer Program.

%\bibliography{mybibfile}

\end{document}